\documentclass[11pt,a4paper]{article}
\pdfoutput=1

\usepackage{jcappub}

\usepackage{printlen}

\usepackage[utf8]{inputenc}
\usepackage{amsfonts}
\usepackage{psfrag}
\usepackage{cancel}
\usepackage{xr}
\usepackage{cleveref}

\crefname{equation}{equation}{equations}
\crefname{figure}{figure}{figures}

\usepackage{verbatim}
\usepackage{multirow}
\usepackage{color}

\usepackage{aas_macros}

\newcommand{\mcomma}{\ensuremath{\;,}}
\newcommand{\mperiod}{\ensuremath{\;.}}

\newcommand{\rhocentral}{\ensuremath{\rho_{\text{bg}}}}

\newcommand{\mcentral}{\ensuremath{M_{\text{bg}}}}
\newcommand{\msol}{\ensuremath{M_{\text{sol}}}}

\newcommand{\Phicentral}{\ensuremath{\Phi_{\text{bg}}}}

\definecolor{ForestGreen}{rgb}{0.3,0.7,0.3}

\definecolor{Purple}{rgb}{0.7,0.0,0.7}

\newcommand{\msolar}{\ensuremath{M_{\odot}}}

\newcommand{\tdecay}{\ensuremath{t_{\text{decay}}}}

\newcommand{\lbox}{\ensuremath{l_{\text{box}}}}

\title{Scalar dark matter vortex stabilization with black holes}

\author[a]{Noah Glennon,}
\emailAdd{nglennon@wildcats.unh.edu}
\affiliation[a]{%
    Department of Physics \& Astronomy
    \\
    University of New Hampshire
    \\
    Durham, NH 03824, USA
}

\author[b]{Anthony E. Mirasola,}
\emailAdd{aem8@illinois.edu}
\affiliation[b]{%
    Department of Physics
    \\
    University of Illinois at Urbana-Champaign
    \\
    Urbana, IL 6180, USA
}

\author[a]{Nathan Musoke,}
\emailAdd{nathan.musoke@unh.edu}

\author[c]{Mark C. Neyrinck,}
\emailAdd{Mark.Neyrinck@gmail.com}

\affiliation[c]{%
    Ikerbasque, the Basque Foundation for Science
    \\
    48009, Bilbao, Spain
}

\author[a]{Chanda Prescod-Weinstein}
\emailAdd{chanda.prescod-weinstein@unh.edu}

\date{\today}

\abstract{%
Galaxies and their dark-matter halos are commonly presupposed to spin. But it is an open question how this spin manifests in halos and soliton cores made of scalar dark matter (SDM, including fuzzy/wave/ultralight-axion dark matter). One way spin could manifest in a necessarily irrotational SDM velocity field is with a vortex. But recent results have cast doubt on this scenario, finding that vortices are generally unstable except with substantial repulsive self-interaction. In this paper, we introduce an alternative route to stability: in both (non-relativistic) analytic calculations and simulations, a black hole or other central mass at least as massive as a soliton can stabilize a vortex within it. This conclusion may also apply to AU-scale halos bound to the sun and stellar-mass-scale Bose stars.
}


\arxivnumber{2301.13220}

\begin{document}

\maketitle

\section{Introduction}

Scalar dark matter (SDM, closely related to, or also known as superfluid, wave, and fuzzy dark matter) models have seen substantial recent attention as a promising alternative to the longstanding WIMP cold dark matter (WIMP CDM) model. Simulations of SDM show that these alternative models behave similarly to WIMP CDM in large-scale structure formation (where WIMP CDM has been successful), but ultralight scalars radically differ on galactic scales. WIMP CDM encounters well-known issues on these scales; there is an apparent deficit of observed satellite galaxies compared to simple estimates from $N$-body simulations of WIMPs, and there is a disagreement between simulated and observed density profiles in galactic cores~\cite{Weinberg2015,Moore1994,Papastergis2015}. Some work has suggested that baryonic effects could possibly account for these discrepancies~\cite{Kim2018,Governato2010}, but this motivation for SDM or warm dark matter remains.
There also are other motivations for SDM, beyond addressing observational problems such as the missing-satellite problem. They are well-motivated in string theory~\cite{Svrcek2006,Arvanitaki2010axiverse,Marsh2016,Hui2017}.

A crucial piece of SDM phenomenology, that distinguishes it from models such as warm dark matter, is in the behavior of its velocity field, which is irrotational, except at vortices. This is because its velocity is defined as a gradient, with zero curl. It is thus of interest whether this picture is consistent with our understanding and observations of spin in the Universe.
Astronomical objects such as galaxies and stars in the Universe generally have some nonzero angular momentum, whether they are made of dark matter or baryons. Intergalactic-scale torques should spin up both baryons and dark matter similarly, e.g.\ with similar angular momentum per unit mass.  One way to understand this is that the torquing is gravitational, which by the equivalence principle should affect all matter equivalently. If an object's angular momentum is drawn from some continuous distribution about zero, it has formally zero chance of being exactly zero. 
this spin presents a question in a SDM scenario. How can an object that is part of an irrotational flow be said to spin?

It is not trivial to explain cosmological-scale spin even in galaxies, because the primordial velocity field is thought to have had negligible vorticity since it stretched away during and after inflation, leaving only a gravity-sourced potential flow until collapsed (multistream) objects form. These then have angular momenta distributed with some nonzero width around zero. The standard, widely accepted explanation for how a cosmological-scale object comes to rotate as it forms in a primordially irrotational velocity field is the tidal torque theory~\cite{Peebles1969}: protogalaxies and protohalos, the regions of the primordial, homogeneous density field that collapse to form galaxies and the dark-matter halos around them, are never perfectly spherical. Their aspherical protuberances are generally torqued up by gravitational tidal fields. This can also be understood as the protuberances carrying some nonzero primordial gravity-sourced velocities, whose contribution to the final collapsed object's angular momentum does not cancel out~\cite{NeyrinckEtal2020}.

Coming back to the question of spin in SDM, there are three mechanisms for a patch of an irrotational fluid such as SDM to carry angular momentum. First, it can sport vortices, where the density goes to zero, and the vorticity to infinity. This mechanism is unusual from an astrophysical standpoint, but it is seen in laboratory superfluid-torquing experiments (e.g.~\cite{MatthewsEtal1999}). A second mechanism to exhibit angular momentum in SDM is in an inhomogeneous density field. If the density is spherically symmetric, the angular momentum will be zero within a sphere. But if we increase the density in a moving lump near the edge of the sphere, that will generally unbalance the angular momentum integral and make the total nonzero. A third mechanism, possible even in a homogeneous density field, is if the patch has an aspherical boundary. In this case, regions outside a maximal sphere about the center generally will contribute angular momentum (e.g.~\cite{NeyrinckEtal2020}). An example of such an object is a Riemann S-ellipsoid~\cite{RindlerDaller2012}. Torqued-up SDM subhalos in cosmological simulations have been observed to resemble these ellipsoids~\cite{DuEtal2018}.

While these latter two mechanisms exist, we imagine them not to be able to carry arbitrarily large spin except in contrived cases. We are most interested in the first, i.e.\ `spin-driven vortices' that would be produced when an object is sufficiently spun up. They are particularly interesting because of the exotic phenomenology, that has the most plausible observational relevance. Spin-driven vortices contrast conceptually with `random vortices,' meandering loci where the wave function happens to go precisely to zero; these are straightforward to study with random wave interference~\cite{HuiEtal2021}. In an interfering jumble far from the center of a dark-matter halo, it is not completely clear how to distinguish random from spin-driven vortices apart from their different formation mechanism, i.e., from a single snapshot of the wave-function. But if the vortex threads a soliton core, another piece of SDM phenomenology, it would be hard not to identify it with physical spin.

Solitons are stable structures thought to reside in the centers of SDM halos, from both simulations and theoretical arguments~\cite{Schive:2014dra,Guth2014,Chavanis2020,Schwabe2016,MoczEtal2017,MoczEtal2019}. They are supported against gravitational collapse by ``quantum'' pressure~\cite{Ruffini1969,Hu2000}, and in some models, by repulsive self-interactions~\cite{Reig2019} so long as certain soliton mass limits are not exceeded (in the case of attractive self-interactions)~\cite{Chavanis2011Analytics,Chavanis2011Numerics,Chavanis2019}. These structures are long-lived, localized solutions which are stable against perturbations, and so are commonly referred to as solitons.
These solitons are expected to form through gravitational thermalization in the centers of SDM halos, and they have a variety of observational effects which could lead to their identification.

It would be a major change in the standard SDM soliton picture if vortices commonly inhabited solitons. Some previous studies have argued that this arrangement is unstable (and therefore uncommon) unless the SDM has sufficiently strong repulsive self-interactions~\cite{Dmitriev:2021utv,Schobesberger2021}. That is, a soliton with one vortex inside the core transitions to a system with the same angular momentum and many vortices outside the soliton core. While vortices still exist in such a system, when they are far away from the soliton core, they exist in regions of much lower density, or can become lost in the chaotic halo that surrounds the soliton. Thus for all practical purposes the vortex has decayed out of the system.

In this work, we show that there is another scenario besides strongly repulsive self-interactions that can stabilize vortices within soliton cores: a background gravitational well generated not by the SDM itself, but by other matter.
While this does not change the energy analysis, and the rotating soliton still is not the energy minimum with fixed mass and angular momentum, it does suppress the decay channel, causing the rotating soliton to have an extremely long lifetime.
When the soliton lives in such a gravitational potential sourced by other matter (baryonic, black-hole, or ambient non-solitonic dark matter), the vortex can have a long lifetime before decaying. This stabilizing mechanism is relevant because all SDM solitons are expected to form in galactic cores where there is a concentration of other matter. Particularly, supermassive black holes (mass $10^6$ to $10^9 \;\msolar$,) are thought to be present in the centers of almost all galaxies  (e.g.~\cite{MagorrianEtal1998}), plausibly coincident with soliton centers. Simulations exist that include SDM, hydrodynamics, and star formation~\cite{MoczEtal2019,KulkarniEtal2022}, but none as far as we know that self-consistently include central black holes as well. Thus, even if the situation of a vortex-soliton supported by a supermassive black hole were common, we would not expect to find them in previous cosmological simulations.

Interactions between the angular momentum of supermassive black holes and their environments is important, for example to the merger of such black holes~\cite{Milosavljevic:2002ht}.
The literature contains other studies of the interaction between scalar and ultralight dark matter and black holes, including: dynamical friction due to black holes moving through scalar dark matter fields~\cite{Wang:2021udl,Vicente:2022ivh,Boudon:2022dxi}; the formation of black holes in scalar dark matter halos~\cite{Escorihuela-Tomas:2017uac,Avilez:2017jql}; the formation of vortex-less solitons through accretion of dark matter on black holes and interpretation as black-hole hair~\cite{Vieira:2014waa,Clough:2018exo,Boskovic:2018rub,Clough:2019jpm,Hui:2019aqm,Brax:2019npi,Jusufi:2020cpn,Bamber:2020bpu,Yuan:2022nmu,Barranco:2011eyw,Barranco:2012qs,Herdeiro:2014goa}; black-hole superradiance and interactions between spinning black holes and scalar halos~\cite{Arvanitaki:2010sy,Cardoso:2011xi,Brito:2015oca,Ferreira:2017pth,Cardoso:2018tly,Baryakhtar:2020gao,Sanchis-Gual:2020mzb,Bamber:2020bpu,Delgado:2020hwr,Delgado:2022pwo}, and the merger of black holes in wave dark matter environments~\cite{Ikeda:2020xvt,Baumann:2022pkl,Bamber:2022pbs}.
The last two classes are particularly pertinent; they are scenarios with significant angular momentum.

Another, recently found example of cosmological-scale rotation is in filaments (comprised of gas, dark matter, and small galaxies) that generally connect neighboring galaxies~\cite{XiaEtal2021,WangEtal2021}. Simulations in a SDM scenario that include hydrodynamics~\cite{MoczEtal2019} have found that intergalactic filaments often contain soliton tubes in their centers, although these seem to fragment into cores on long timescales. It would be quite interesting if spin-driven vortex tubes often resided in filaments. But in a 2D version of the 3D discussion above about halo spin, an irrotational SDM velocity field could carry some amount of filamentary angular momentum without vortex tubes, through velocity or density asymmetry in the filament's cross section.

The route to stability we find in this paper does not immediately pertain to vortex tubes within intergalactic filaments, because there is no known stringlike analog of a black hole that would gravitationally support a vortex tube. Still, it would be interesting to further investigate in simulations what form filament spin takes in SDM, and if vortex tubes might often inhabit filaments. It is possible that they could have long decay/fragmentation lifetimes, as the soliton tubes themselves have, or could be supported by repulsive self-interactions.

This paper is organised as follows.
In \cref{sec:theory} we analyse theoretical considerations for rotating solitons in the presence of a background potential.
In \cref{sec:solution} we present an approximate solution for vortices stabilised by the gravitational potential of a point mass.
In \cref{sec:instability-mode} we discuss the energy and decay modes of a rotating soliton in an central gravitational potential, making analytic arguments that support the results found in simulations.
In \cref{sec:simulations}, we use simulations from \texttt{UltraDark.jl} to demonstrate that vortex solitons can be long-lived even in the presence of perturbations.
We end with a discussion of possible applications and future directions in \cref{sec:discussion}.

\section{Analytic description of vortices in SDM}
\label{sec:theory}

\subsection{Rotating solutions to Gross--Pitaevskii--Poisson equations}
\label{sec:solution}

We are interested in scalar particles with mass $m_a$ and Lagrangian density
\begin{equation}
    \label{eq:lagrangian}
    \mathcal{L}
    =
        \frac{1}{2} g^{\mu\nu} \partial_{\mu} \phi \partial_{\nu} \phi
        - \frac{1}{2} m_a^2 \phi^2
        - \frac{\lambda}{4} \phi^4
    \mperiod
\end{equation}
In the non-relativistic limit, the field $\phi$ can be rewritten as
\begin{equation}
    \phi =
    \frac{1}{\sqrt{2m}}
    \left(
    e^{-imt} \psi + e^{+imt} \psi^*
    \right)
    \mcomma
\end{equation}
and transformed to a field $\psi$ whose equations of motion are the Gross--Pitaevskii--Poisson equations
\begin{gather}
    \label{eq:schrodinger}
    i \frac{\partial \psi}{\partial t}
    =
    - \frac{1}{2m} \nabla^2 \psi
    + m \psi \left(\Phi + \Phicentral \right)
    + \frac{\lambda}{2 m^2} |\psi|^2 \psi
    \\
    \label{eq:poisson}
    \nabla^2 \Phi(\mathbf{r}) = 4 \pi G \rho(\mathbf{r}) = 4 \pi |\psi|^2
    \mperiod
\end{gather}
Here $\Phi$ is the gravitational potential sourced by the ULDM density $|\psi|^2$, $\Phicentral$ is the gravitational potential due to a mass with density $\rhocentral$, and $\lambda$ parametrizes the self-interactions arising from a quartic term in the SDM Lagrangian, with $\lambda>0$ corresponding to repulsive self-interactions. Such a quartic self-interaction is common in axion models arising from string theory~\cite{Arvanitaki2010axiverse}.
We do not consider backreaction on the background density, instead treating $\rhocentral$ as a fixed background.

The simplest solutions to the Gross--Pitaevskii equations that carry non-zero velocity circulation are axially symmetric solutions with a central vortex line, characterized by circulation $l$, which is an integer defining the winding number of the phase around the vortex line~\cite{Dmitriev:2021utv}. These solutions have the form
\begin{equation}
    \psi_l (r,z) e^{-i \omega_l t + i l \phi},
\end{equation}
where a vortex line of circulation $l$ is located at $r=0$.
The phase-independent amplitude $\psi_l (r,z)$ can be explicitly written if self-interactions vanish ($\lambda = 0$) and the background potential in \cref{eq:schrodinger} is due to a central point mass with $\Phicentral \gg |\Phi|^2$.
Then \cref{eq:schrodinger} is approximately
\begin{gather}
    \label{eq:schrodinger-point-mass}
    i \frac{\partial \psi}{\partial t} \approx - \frac{1}{2m} \nabla^2 \psi - m \psi \frac{\mcentral}{r}
    \mperiod
\end{gather}
The gravitational potential from the point mass is analogous to the Coulomb potential, so in this limit, the rotating soliton is well-described by hydrogen atom wavefunctions~\cite{Silveira1995,Baumann:2019eav,Santos:2020sut,Guo:2020}. 
The lowest energy solution exhibiting a vortex with circulation 1 is
\begin{equation}
    \label{eq:h-ansatz}
    \psi_{2, 1, \pm 1}
    =
    \frac{1}{2 \sqrt{6}} a^{-3/2} \left(\frac{r}{a}\right) e^{-r/2a}
    \left(- {\left(\frac{3}{8\pi}\right)}^{1/2} \sin\theta e^{\pm i \phi}\right)
\end{equation}
where $a = 1/\mcentral$, $r$ is the radial coordinate, $\theta$ is the polar angle, and $\phi$ is the azimuthal angle. 
Due to the nonlinearities in the Gross--Pitaevskii--Poisson equations, the exact solution exhibits deviations from the hydrogen wavefunctions in its radial profile. 
Indeed, we observe in simulations that starting from a radial profile that deviates from that of the ground state results in radial-profile oscillations, but not azimuthal perturbations. In the absence of a background potential, there are significant deviations from these hydrogen-like solutions. However, qualitative properties such as axial symmetry, and the presence of a central vortex in the middle of the densest region of the soliton remain~\cite{Kling2021}. 

We refer to all such systems as ``vortex solitons'' to indicate the essential point that the vortex is located in the center of the soliton, although we note that these objects may not satisfy all properties normally associated with solitons, in particular, their stability against collisions and perturbations.

When the vortex line passes through the soliton core, as it does in these hydrogen-like wavefunctions, the angular momentum $L_{tot}$ is quantized to integer multiples of the particle number $N$ in the soliton, $L_{tot} = \hbar N l$. However, the system can carry a lower value of angular momentum if the vortex line departs from the center of the soliton.
In the limit that the vortex is far from the soliton center, the angular momentum asymptotically approaches zero.
When the vortex is far away from the soliton, there is hardly any mass near the vortex. The mass, and hence angular momentum density, is concentrated in the soliton core. But the velocity field is highest near the vortex, because the velocity is proportional to the phase gradient, and the velocity falls off as one over the distance from the vortex. Therefore when the vortex is displaced from the soliton core, the total angular momentum can take any value $L<L_{tot}=Nl$.

\subsection{Instability of vortex-solitons}
\label{sec:instability-mode}

Ref.~\cite{Dmitriev:2021utv} demonstrated that in the absence of repulsive self-interactions, the vortices in rotating solitons of scalar field dark matter are unstable. They showed that the vortex state is not the lowest energy state; it can attain lower energy by shedding angular momentum.
They did not consider the influence of a central background potential. We show first that when an background potential $\Phi_\mathrm{bg}$ caused by point mass $M_\mathrm{bg}$ is included, the vortex state is still not the lowest energy state, so it cannot truly be stable. We then show that the decay channel is suppressed, leading to the long lifetimes of the vortices.

Begin with an initial state in the configuration $\psi' _{s}$, with circulation $l>0$ and particle number $N_s$. We will construct a configuration with the same particle number $N_s$ and total angular momentum $\hbar l N_s$, but lower energy, and without a vortex in the soliton core. This shows that the vortex is not a global minimum energy state in the sector with angular momentum $l$, so it is not stable to sufficiently large perturbations.

To construct the lower-energy configuration, we remove $dN_s$ particles from the initial state, and add $dN_0$ particles to the $l=0$ mode and $dN_{l'}$ to the $l'\gg l$ mode. In order to conserve particle number, we require
\begin{equation}
    dN_s = dN_0 + dN_{l'}.
\end{equation}
In order to conserve angular momentum, we require
\begin{equation}
    l dN_s = l' dN_{l'}.
    \label{eq:Lconservation}
\end{equation}
Thus the final wavefunction is
\begin{equation}
    \psi' _s \rightarrow \psi = \psi_s + \sqrt{dN_0} \Psi_0 + \sqrt{dN_{l'}} \Psi_{l'},
\end{equation}
where the magnitude of the original soliton configuration has decreased, $|\psi_s(\mathbf x)|<|\psi' _s(\mathbf x)|$. 

We will show that the energy of the initial rotating soliton is greater than the energy of this new configuration that has the same angular momentum.
The energy of the final configuration is
\begin{equation}
    E_f = E_s + \omega_0 dN_0 + \omega_{l'}d N_{l'},
\end{equation}
where $\omega_{l'}$ is the energy of a single particle in state $l'$, and $E_s$ is the energy of all particles remaining in the initial state of the rotating soliton.
The initial energy is
\begin{equation}
    E_i = E_s +\omega_s dN_s + O(dN_s ^2),
\end{equation}
so the change in energy is
\begin{equation}
    \Delta E = (\omega_0-\omega_s)d N_0 +(\omega_{l'}-\omega_s)d N_{l'}.
    \label{eq:DeltaE}
\end{equation}
To show that this configuration has a lower energy than the vortex soliton, we must estimate the energy differences $\omega_{l'}-\omega_s$, for $l'=0$, and for $l'\gg 1$.

The modes $\Psi_{l'}$ and energies $\omega_{l'}$, are the solutions to the non-interacting Gross--Pitaevskii equation with background potential generated by the soliton and (in our case) the background central potential:
\begin{equation}
    \omega_{l'}\Psi_{l'} = -\frac{\nabla^2 \Psi_{l'}}{2m}+m(\Phi_s+\Phi_\mathrm{bg})\Psi_{l'}.
\end{equation}
For any $l'<l$, where $l$ is the angular momentum of the soliton, we must have $\omega_{l'}< \omega_s$, because the centrifugal barrier is weaker, while the other energy terms are the same. So for $l=0$, we have
\begin{equation}
\omega_0-\omega_s<-\int d^3 x \frac{l^2 |\Psi_l|^2}{2mr^2}<0.
\label{eq:gap}
\end{equation}
The background potential does not depend on $l$, so it does not affect this estimate of the energy difference. In fact, it further lowers the energy of the $l'=0$ state, because that state has more mass closer to the center of the potential. 

Now consider the $l'\gg 1$ state. The wavefunctions become more and more spatially spread out as $l'$ increases, so we can approximate both the wavefunctions and the energies as the eigenstates of the hydrogen atom with mass $M_s + M_\mathrm{bg}$ (since the soliton behaves like a point mass when viewed from long distances). The energies are therefore
\begin{equation}
    \omega_{l'} \approx -\frac{m^3 G^2 {(M_s + M_\mathrm{bg})}^2}{2{(l'+1)}^2} \sim O({l'} ^{-2}).
\end{equation}
Note that the background potential significantly increases the energy gaps between states with increasing $l'$ (by adding a term proportional to $M_\mathrm{bg} ^2$, but does not affect the scaling of the energies with $l'$). In particular, as $l'\rightarrow \infty$, the energies asymptotically approach zero. 

Returning to Eq. (\ref{eq:DeltaE}), we can re-write the energy difference as
\begin{equation}\begin{aligned}
    \Delta E &= (\omega_0-\omega_s) d N_0 + (O({l'} ^{-2}) -\omega _s) dN_{l'}\\
    &= (\omega_0-\omega_s) d N_0 + (O({l'} ^{-2}) -\omega _s) \frac{l}{l'} dN_s\\
    &= (\omega_0-\omega_s)dN_s + O({l'} ^{-1})d N_s <0,
\end{aligned}\end{equation}
where in the second line we used Eq. (\ref{eq:Lconservation}), which ensures conservation of angular momentum. Therefore, for sufficiently large $l'$, the energy gap is dominated by the first term, which we estimated in Eq. (\ref{eq:gap}) is negative. This shows that the initial vortex state is not a global energy minimum over $l$, so it is not stable against perturbations.

While the vortex state is not a global energy minimum, it might still be extraordinarily long lived. To decay from the initial vortex state into the lower energy configuration constructed above, we would require transfers from the initial $l=1$ state into an $l'\gg1$ state, but these are not strongly coupled together. In simulations of vortex solitons  without background potential (which do show an instability)~\cite{Dmitriev:2021utv}, the decay has been demonstrated to occur due to pairwise transitions from $l$ state to neighboring $l'=l\pm1$ states. In particular, an initial $l=1$ loses occupancy to $l'=0,2$ states, whose occupation numbers grow. At later times, the occupation of the $l=2$ state slows while $l'=1,3$ begins to grow.

The transitions from the initial state into a decay state can only take place when there are finite transition matrix elements between the coupled states, so that there is a decay channel between the coupled states. These transition matrix elements can be estimated in a perturbation theory~\cite{Zagorac:2021qxq}. In the limit where the background mass is significantly greater than the soliton mass, $M_{\text{bg}}/M_s \gg 1$, all transition matrix elements vanish, because the background potential dominates the self-potential of the soliton and the Gross--Pitaevskii--Poisson equations become linear equations with exact eigenstates corresponding to hydrogen atom wavefunctions. Thus in this limit we expect the soliton to be long-lived and stable against all perturbations. However, the transition matrix elements are suppressed to the first order in perturbation theory even when the nonlinearities of the system are still significant. In particular, the decay mode identified by Ref.~\cite{Dmitriev:2021utv} is suppressed by the background potential.

At the first order in perturbation theory, the stationary states are the eigenstates of the background potential. Since our background potential is that of a point mass, these are the hydrogen atom wavefunctions. The self-gravity of the soliton creates a perturbative potential $\Delta V$ that introduces transition matrix elements between the unperturbed stationary states. In our initial state in Eq. (\ref{eq:h-ansatz}), the mass distribution is axially symmetric, $\rho(r,\theta,\phi)=\rho(r,\theta)$, and hence the gravitational potential is axially symmetric as well. In this case, the mass distribution and gravitational potential admits a multipole expansion in terms of Legendre polynomials $P_n(\cos\theta)$,
\begin{align}
    \rho(r,\theta) &= \sum_n \rho_n(r) P_n (\cos\theta)\\
    \Phi(r,\theta) &= \sum_n \Phi_n(r) P_n (\cos\theta),
\end{align}
where the components $\Phi_n$ of the potential are determined by
\begin{equation}
    \Phi_n(r) = -\frac{2\pi G}{n+\frac{1}{2}} r^{-n-1} \int_0 ^r {r'} ^{n+2} \rho_n(r') dr' - \frac{2\pi G}{n+\frac{1}{2}} r^n \int_r ^\infty {r'} ^{1-n} \rho_n (r') dr'.
\end{equation}

In our case, the initial state has a mass distribution with only $P_0$ and $P_2$ components nonzero. Therefore these are the only nonvanishing components of the potential caused by the rotating soliton. Moreover, the background gravitational potential caused by the point mass is a central potential and so only has nonvanishing $P_0$ component. These monopole moments conserve angular momentum and do not lead to any transitions between states of different $l$. The component proportional to $P_2$ does allow transitions. The transitional matrix element is 
\begin{equation}
    \langle l' m' | \Delta V | l m \rangle = \int f'^*(r) \Phi_2(r) f(r) r^2 dr \int {Y_{l'} ^{m'}} ^* Y_2 ^0 Y_l ^m d\Omega,
\end{equation}
where $f,f'$ are the radial wavefunctions of the initial and final states and $Y_l ^m$ are spherical harmonics.
The integral over angles is a Wigner 3-j symbol,
\begin{equation}
\langle l' m' | \Phi | l m \rangle \propto 
\begin{pmatrix} l &2 &l' \\ m &0 &-m'\end{pmatrix}.
\end{equation}
The Wigner 3-j symbols are nonzero only when their selection rules are satisfied. For our initial rotating soliton state $|l=1,m=1\rangle$, we have permitted transitions to $|l=2,m=1\rangle$, $|l=3,m=1\rangle$, and no others. In particular, the pairwise transition which leads to the decay of the rotating soliton without background potential is \emph{not} permitted. Thus the dominant instability mode observed by Ref.~\cite{Dmitriev:2021utv} is not initially present in this system, and is suppressed by the background potential, at least at first order in perturbation theory. This suppression of the dominant instability mode of the soliton without background potential leads to the long lifetime of our vortex solitons compared to those in other studies.

\section{Simulations of vortex solitons}
\label{sec:simulations}

We use \texttt{UltraDark.jl} to simulate these approximate solutions and understand their dynamics and stability.
\texttt{UltraDark.jl} is a pseudospectral solver for the Gross--Pitaevskii--Poisson equations that allows for contributions from background gravitational fields, such as the potential $\Phicentral$ in \cref{eq:poisson}~\cite{ultradark}.

\texttt{UltraDark.jl} uses code units for time, length and mass,
\begin{gather}
    \label{eq:time_scale}
    \mathcal{T}
    =
    {\left(
        \frac{3}{8 \pi} H_0^2 \Omega_{m, 0}
    \right)}^{-1/2}
    \approx
    74 \;\mathrm{~Gyr}
    \\
    \label{eq:length_scale}
    \mathcal{L}
    =
    {\left(
            \frac{\hbar}{m}
    \right)}^{1/2}
    {\left(
            \frac{3}{8\pi}
            \Omega_{m, 0} H_0^2
    \right)}^{-1/4}
    \approx
    38 {\left(\frac{10^{-22}\mathrm{~eV}}{m}\right)}^{1/2} \;\mathrm{~kpc}
    \\
    \label{eq:mass_scale}
    \mathcal{M}
    =
    {\left(
            \frac{\hbar}{m}
    \right)}^{3/2}
    \frac{1}{G}
    {\left(
            \frac{3}{8\pi} \Omega_{m, 0} H_0^2
    \right)}^{1/4}
    \approx
    2.2 \times 10^{6} {\left(\frac{10^{-22}\mathrm{~eV}}{m}\right)}^{3/2} \;\msolar
    \mperiod
\end{gather}
In these units, which we write as primed, \cref{eq:schrodinger,eq:poisson} become
\begin{gather}
    i \frac{\partial\psi'}{\partial t'}
    =
    - \frac{1}{a^{2}} \nabla'^{2} \psi'
    + \psi' \left(\Phi'(\mathbf{r}') + \Phicentral'(\mathbf{r}) \right)
    \\
    \nabla'^{2} \Phi'(\mathbf{r})
    =
    \frac{1}{a} 4 \pi |\psi'|^{2}
    \mperiod
\end{gather}
We use $H_0 = 70 \;\mathrm{~km/s/Mpc}$ and $\Omega_{m,0} = 0.3$.
In these units, the critical density of the Universe today is $1 \;\mathcal{M}/\mathcal{L}^3$.
These units are dependent on the mass $m$ of SDM particle considered, and so the following simulations can be interpreted as corresponding to different scenarios for different values of $m$.
In the interest of clarity, units in figures assume $m = 10^{-22}$, with the understanding that a different choice of $m$ would scale the units as above.
We run simulations with a resolution of $256^3$ and a box length of $\lbox = 22/\mcentral$ set by the length scale in the initial wavefunction \cref{eq:h-ansatz}.

\texttt{UltraDark.jl} has periodic boundary conditions.
The density at the edge of the grid is $\lesssim 1\%$ of the maximum density, and so only a small amount of matter passes through the boundary. 
The periodic boundary conditions also cause the solutions to see gravitational fields due to the scalar field in neighbouring boxes; this adds a violation of spherical symmetry beyond that of the Cartesian grid.
The gravitational potential due to the central mass does not experience these effects.

Our simulations assume Newtonian gravity, but are sufficient to model black holes on these scales.
The characteristic length scale of a vortex-less soliton is~\cite{Guzman:2004wj,Marsh:2015wka,Herrera-Martin:2017cux}
\begin{equation}
     \frac{r_s}{\mathrm{pc}}
     =
     7.6 {\left(\frac{m}{10^{-22} \;\mathrm{eV}}\right)}^{-2} {\left(\frac{\msol}{10^{11} \msolar}\right)}^{-1} \mperiod
\end{equation}
This is roughly the major radius of a vortex soliton.
The ratio of the black hole radius and vortex soliton radius then goes as
\begin{equation}
    \frac{r_s}{r_{\text{Schwarzschild}}}
    \sim
    10 {\left(\frac{m}{10^{-22} \;\mathrm{eV}}\right)}^{-2} {\left(\frac{\msol}{10^{11} \msolar}\right)}^{-1} {\left(\frac{\mcentral}{10^{13} \msolar}\right)}^{-1} \mperiod
\end{equation}
With $m \sim 10^{-22}\;\mathrm{eV}$ and $\msol \sim \mcentral \sim \mathcal{O}(10^{7}) \;\msolar$, this ratio is $r_s/r_{\text{Schwarzschild}} \sim 10^{10}$.
Even for $\msol \sim \mcentral \sim \mathcal{O}(10^{10}) \;\msolar$, there is a large separation in scales with $r_s/r_{\text{Schwarzschild}} \sim 10^{5}$.
This means that although our simulation setup cannot accurately describe cases with sufficiently massive ALPs, solitons, or central masses, those considered here are within this bound.
Furthermore, the Schwarzschild radius of a black hole with $\mcentral = 10^7 \; \msolar$ would be $\lesssim 1 \times 10^{-6} \; \mathrm{kpc}$, far smaller than the grid spacing of $\sim 10^{-1} \; \mathrm{kpc}$.
In addition, the vortex has vanishing density at the center so accretion would be minimal.

\begin{figure}
    \centering
    \includegraphics[width=\linewidth]{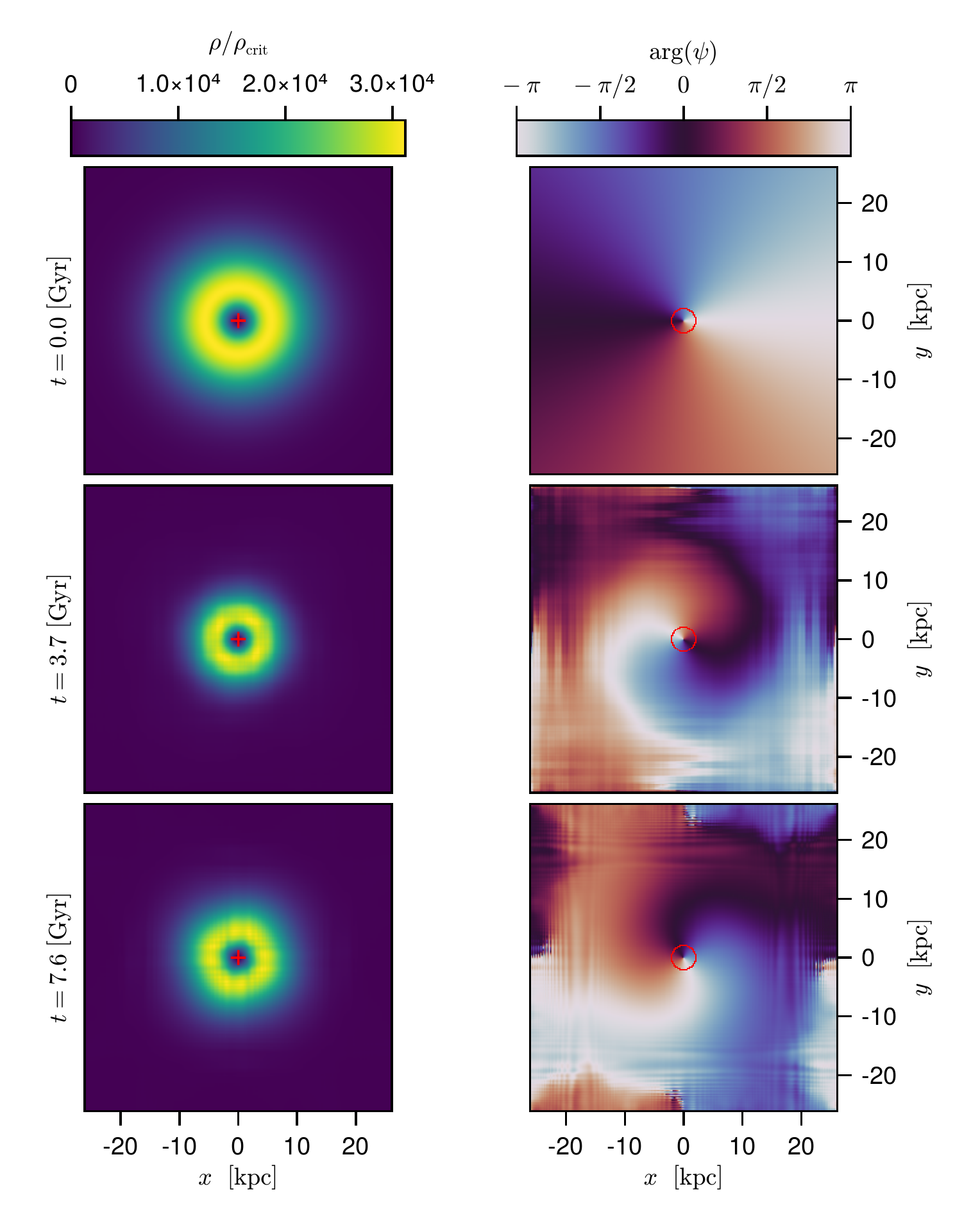}
    \caption{%
        (Stable vortex-soliton, with no initial perturbations.)
        Snapshots of a simulation with initial conditions \cref{eq:h-ansatz} with $\mcentral = 3.5 \times 10^{7} \; M_{\odot} \times {\left(10^{-22} \;\mathrm{eV} /m\right)}^{3/2}$ and $\msol = \sqrt{2} \mcentral$.
        Time increases from top to bottom.
        The $x$- and $y$-grids are the same in each panel.
        The left column shows the density projected into the $x$-$y$ plane.
        The right column is the phase in the $x$-$y$ plane in a slice through $z=0$; the red curve is that used to compute the winding number.
        This initial condition is not an exact equilibrium because $\msol > 0$.
        One can see that although there are radial fluctuations, the vortex persists.
        See \url{https://www.youtube.com/watch?v=dEHL1Io0akY} or \url{https://doi.org/10.5281/zenodo.7675830} for an animation.
    }
    \label{fig:no-perts-stable}
\end{figure}

\begin{figure}
    \centering
    \includegraphics[width=\linewidth]{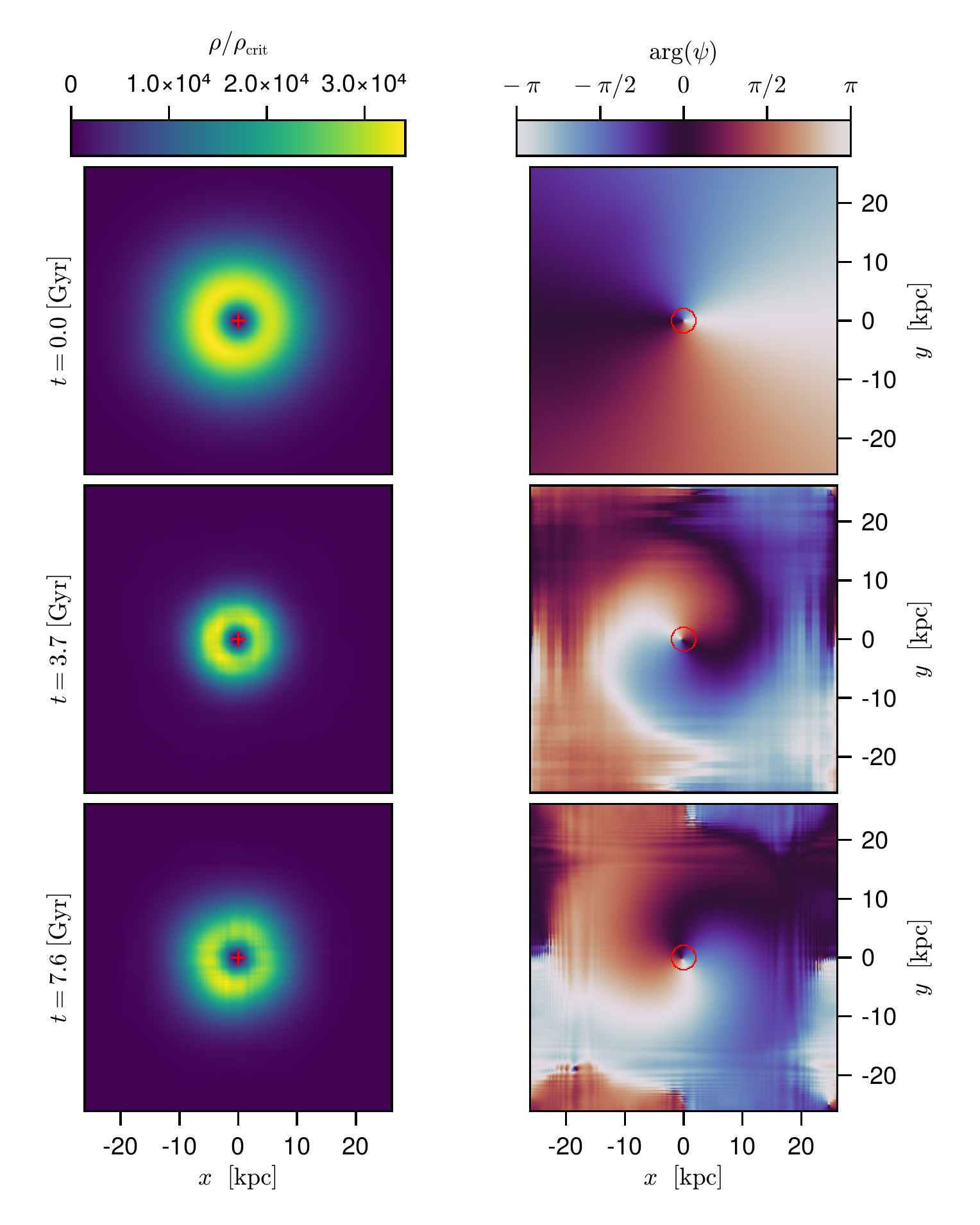}
    \caption{%
        (Stable vortex-soliton, with initial perturbations.)
        Snapshots of a simulation with initial conditions \cref{eq:h-ansatz} with $\mcentral = 3.5 \times 10^{7} \; M_{\odot} \times {\left(10^{-22} \;\mathrm{eV} /m\right)}^{3/2}$, $\msol = \sqrt{2} \mcentral$ and perturbations as in \cref{eq:pert}.
        Time increases from top to bottom.
        The left column is the projected density.
        The right column is the phase in the $x$-$y$ plane in a slice through $z=0$; the red curve is that used to compute the winding number.
        This scenario does not have axial symmetry, but the central vortex persists.
        See \url{https://www.youtube.com/watch?v=DYeL5UHQjdE} or \url{https://doi.org/10.5281/zenodo.7675830} for an animation.
    }
    \label{fig:perts-stable}
\end{figure}

\begin{figure}
    \centering
    \includegraphics[width=\linewidth]{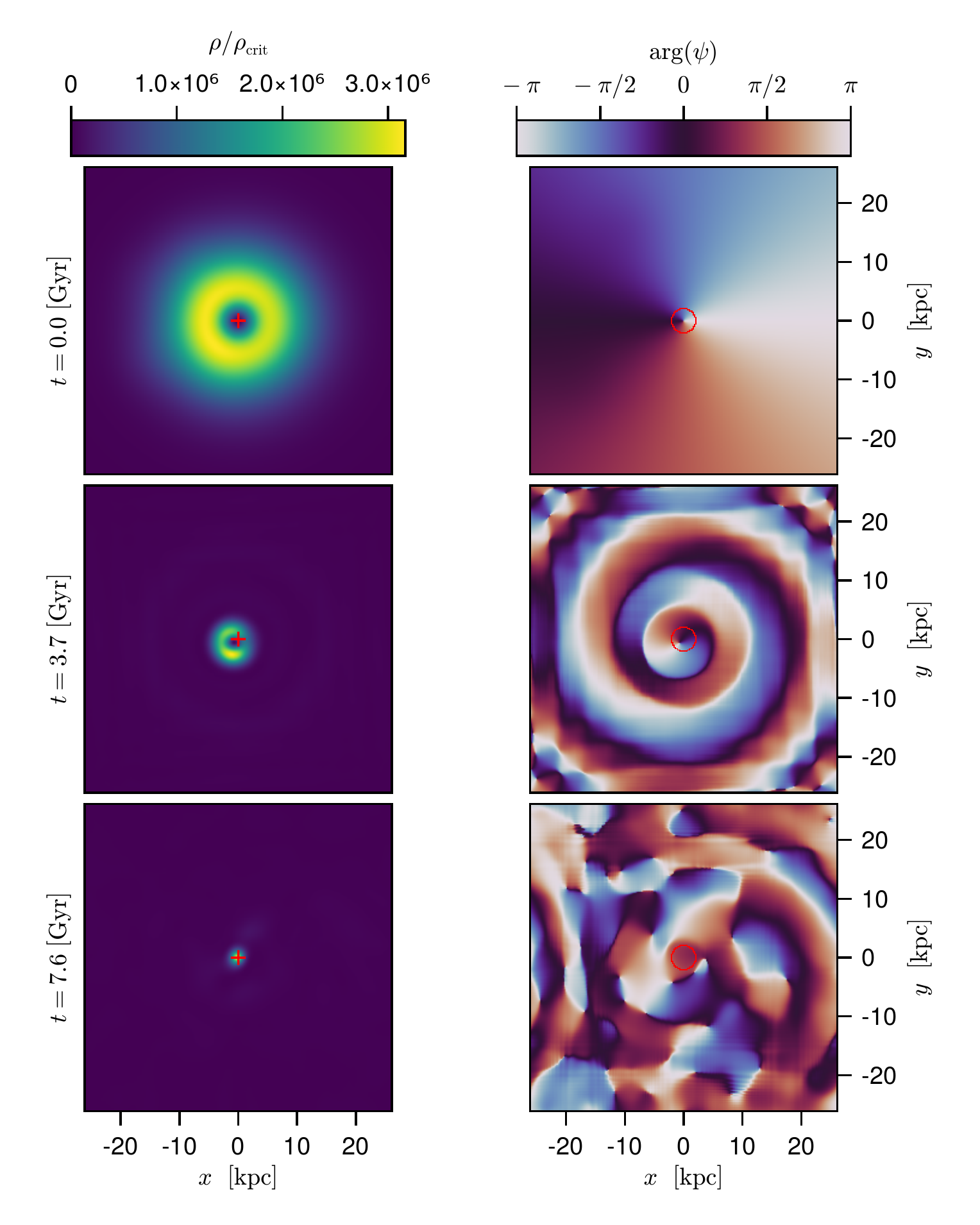}
    \caption{%
        (Unstable vortex-soliton, with initial perturbations.)
        Snapshots of a simulation with initial conditions \cref{eq:h-ansatz} with $\mcentral = 3.5 \times 10^{7} \; M_{\odot} \times {\left(10^{-22} \;\mathrm{eV} /m\right)}^{3/2}$, $\msol = 8 \mcentral$ and perturbations as in \cref{eq:pert}.
        Time increases from top to bottom.
        The left column is the projected density.
        The right column is the phase in the $x$-$y$ plane in a slice through $z=0$; the red curve is that used to compute the winding number.
        The ratio of the soliton mass and central mass is sufficiently large that the vortex is unstable even in the presence of the central mass.
        See \url{https://www.youtube.com/watch?v=g9NVU4LK2Lc} or \url{https://doi.org/10.5281/zenodo.7675830} for an animation.
    }
    \label{fig:perts-unstable}
\end{figure}

We consider a vortex to be stable up to a decay time $t_{\text{decay}}$ if the winding number does not change in this time interval.
We measure the winding number by integrating the phase difference between neighbouring grid points around a loop.
The choice of loop around which to compute the winding number is important but somewhat subjective.
When perturbations are introduced, the vortex orbits the central mass with a small but non-zero radius, even when has not decayed (see for example the second snapshot in \cref{fig:perts-unstable}).
If the loop is too small, the vortex moves outside of it even though it has not decayed.
If the loop is too large, it includes transient vortices in the underdense outskirts of the soliton.
These can increase the winding number if the have the same chirality as the central vortex, or decrease it if they have opposite chirality.
We used loops with radius similar to the radius of a ground state soliton of the same mass.
In practice, this corresponded to a radius of 10 to 12 grid points.
\Cref{fig:no-perts-stable,fig:perts-stable,fig:perts-unstable} show a circle with a radius of 10 grid points.

In \cref{fig:no-perts-stable} we show the results of evolving the initial conditions of \cref{eq:h-ansatz} forward with $\mcentral = 3.5 \times 10^{7} \; M_{\odot} \times {\left(10^{-22} \;\mathrm{eV} /m\right)}^{3/2}$ and $\msol = \sqrt{2} \mcentral$.
One can see that although there are radial oscillations, the vortex persists to the end of the simulation.
There are also perturbations aligned with the simulation grid; these break the  axial symmetry of the ansatz, so should not increase the stability of the vortex.


Verifying stability requires the addition of perturbations that break the cylindrical symmetry of the system.
We perturb the simulation with $10$ small Gaussian overdensities with random position and velocity.
In particular,
\begin{equation}
    \label{eq:pert}
    \delta \psi
    =
    A
    \sum_{j=1}^{N_{\text{pert}}}
    \rho_c
    \exp\left(-\frac{{(\mathbf{r}_j - \mathbf{r})}^2}{\sigma}\right)
    \exp\left(i (\mathbf{r}_j - \mathbf{r}) \cdot \mathbf{v}_j\right)
    \mcomma
\end{equation}
where $\sigma = \lbox / 50$, and for each perturbation the spherical radius $r_j$ is uniformly sampled from the interval $[\lbox/2 \time 0.5, \lbox/2 \times 0.9]$, the azimuthal angle from $[0, 2 \pi]$ and the polar angle from $[0, \pi]$.
Each perturbation is assumed to have an angular velocity $\boldsymbol{\omega}_j$, where each component of $\boldsymbol{\omega}_j$ is drawn from a normal distribution with mean $0$ and standard deviation $1$; $\mathbf{v}_j = \mathbf{r}_j \times \boldsymbol{\omega}_j$.
The same positions and velocities of random perturbations are used for all simulations.
The overall scaling $A$ is set such that the ratio of the mass of the perturbations and soliton is $|\delta \psi|^2 / |\psi|^2 = 1/100$.
A realistic galaxy core might be subject to much larger perturbations, but preliminary simulations suggest that a central mass prolongs the lifetime of vortex-solitons even in the presence of much larger perturbations.
We have not yet fully investigated this area of parameter space.

\Cref{fig:perts-stable} shows the result of adding this perturbation to the approximate solution in \cref{eq:h-ansatz} and evolving it forward.
The other initial conditions are the same as in \cref{fig:no-perts-stable}: $\mcentral = 3.5 \times 10^{7} \; M_{\odot} \times {\left(10^{-22} \;\mathrm{eV} /m\right)}^{3/2}$ and $\msol = \sqrt{2} \mcentral$.
One may note that perturbations are not obvious in the initial phase; this is because the perturbation densities and velocities are small, and many of the perturbations lie outside the plane.
The vortex persists until the end of the simulation, $7.6 \; \mathrm{Gyr}$.

\Cref{fig:perts-unstable} shows snapshots of a scenario which is unstable because the central mass is not dominant.
As in \cref{fig:no-perts-stable,fig:perts-stable}, $\mcentral = 3.5 \times 10^{7} \; M_{\odot} \times {\left(10^{-22} \;\mathrm{eV} /m\right)}^{3/2}$.
However, the vortex-soliton is much more massive, with $\msol = 8 \mcentral$.
As the simulation proceeds, the initial dark matter distribution contracts to form a genus 0 soliton.
This soliton falls to the bottom of the central potential.
Even before the initial central vortex dissipates, it orbits the central potential, instead of remaining centred.

\begin{figure}
    \centering
    \includegraphics[width=0.8\linewidth]{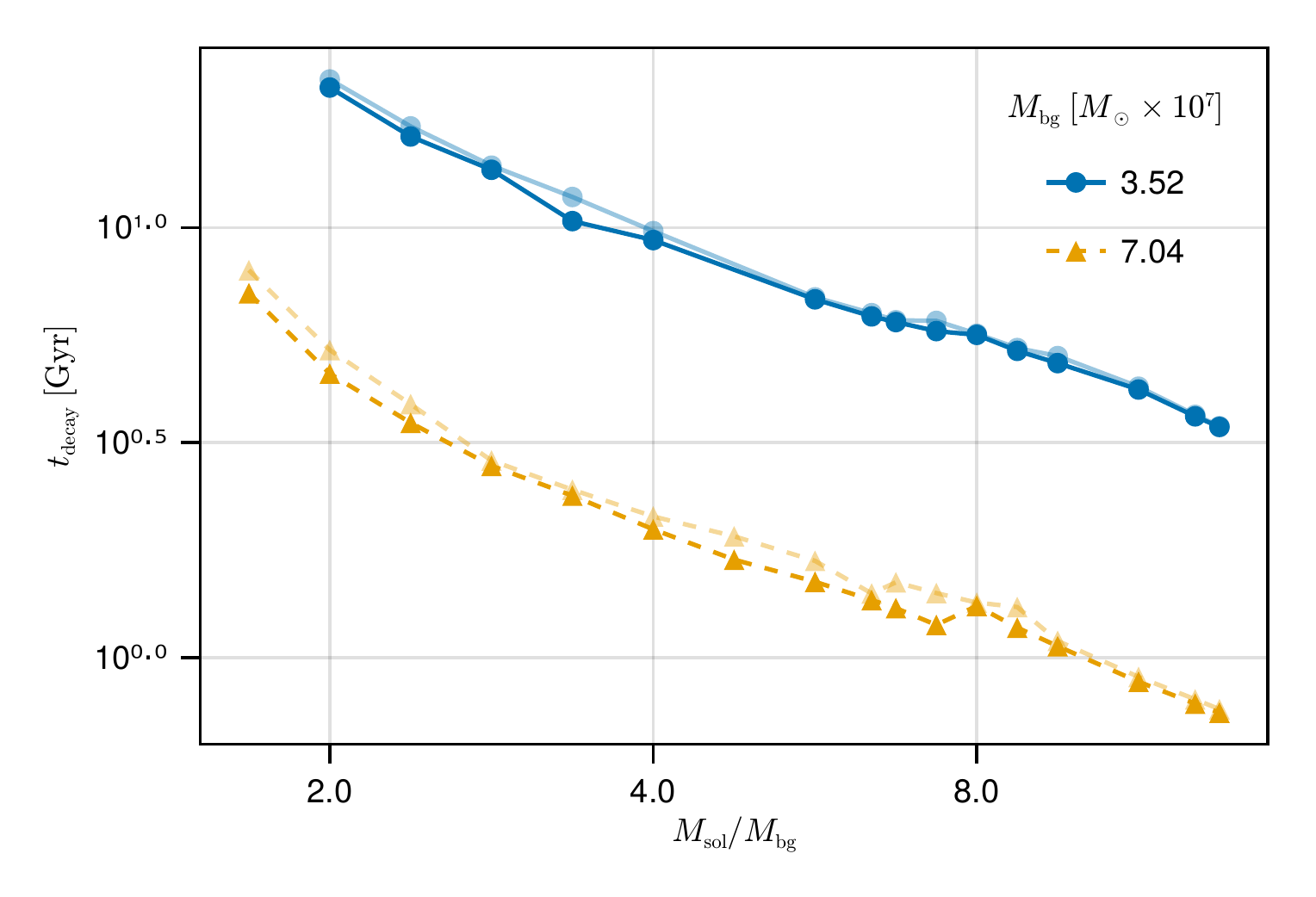}
    \caption{%
        Plot of decay time against the mass ratio $\msol/\mcentral$.
        The solid blue curve with circles has a central mass of $\mcentral = 3.5 \times 10^7 \;M_{\odot} \times {\left(10^{-22} \;\mathrm{eV}/m\right)}^{3/2}$.
        The dashed orange curve with triangles has a $\mcentral = 7.0 \times 10^7 \;M_{\odot} \times {\left(10^{-22} \;\mathrm{eV}/m\right)}^{3/2}$.
        In each case, the bold (faint) curve has winding number computed with an aperture of $10$ ($12$) grid points.
        In both cases, increasing the mass of the soliton relative to the central mass decreases the time taken for the vortex to decay.
    }
    \label{fig:t_decay_vs_m_ratio}
\end{figure}

\begin{figure}
    \centering
    \includegraphics[width=0.8\linewidth]{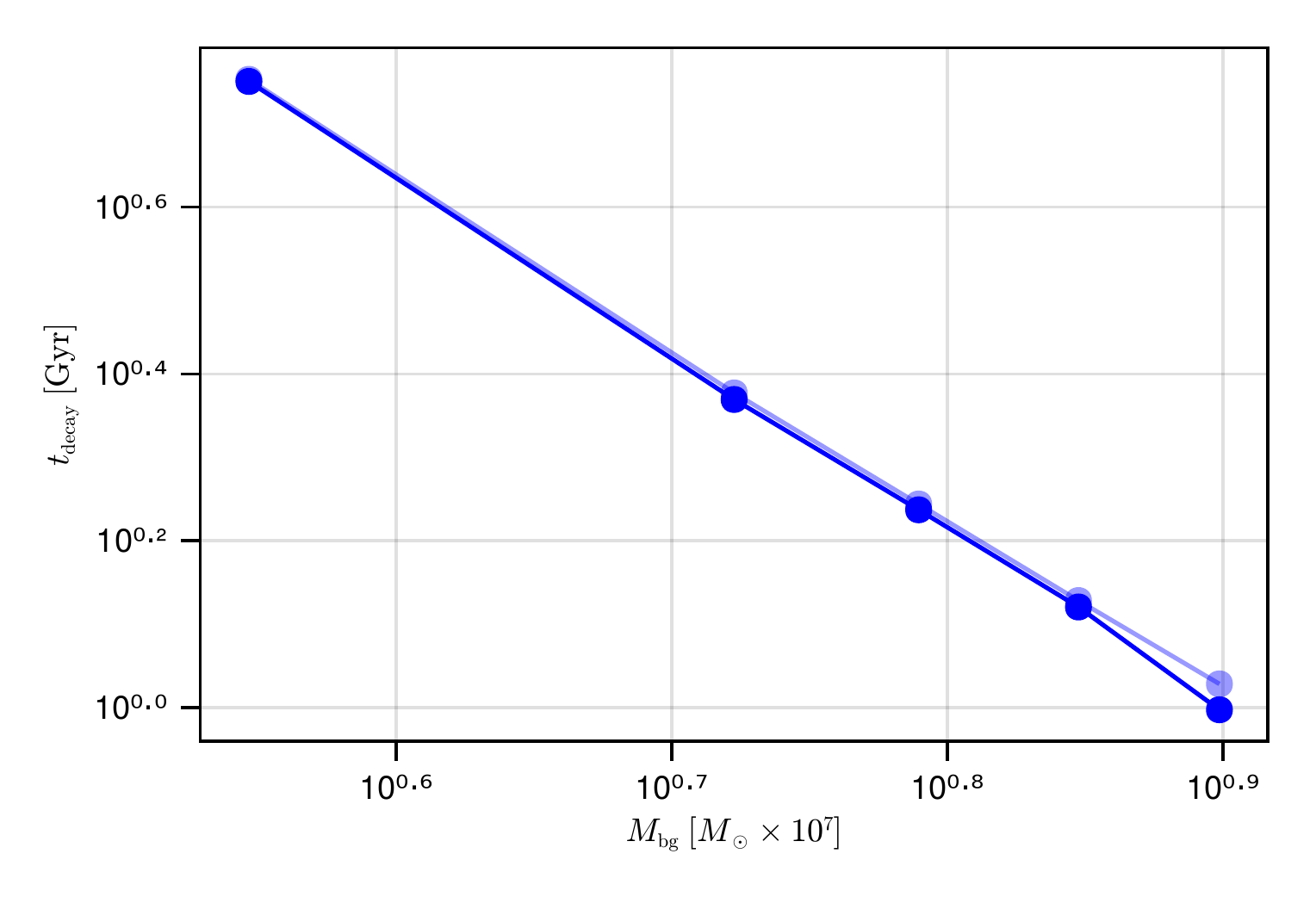}
    \caption{%
        Plot of decay time against the central mass $\mcentral$.
        In each case, $\msol/\mcentral = 8$.
        The bold (faint) curve has winding number computed with an aperture of $10$ ($12$) grid points.
        Increasing the central mass also decreases the stability time at fixed mass ratio.
    }
    \label{fig:t_decay_vs_m_central}
\end{figure}

To explore the relation between the mass ratio $\msol/\mcentral$ and decay time, we ran a set of simulations with $\mcentral=3.5\times 10^7 \;\msolar$ and $\mcentral=7.0\times 10^7 \;\msolar$, and varying $\msol/\mcentral$; the results of this are shown in \cref{fig:t_decay_vs_m_ratio}.
In the interest of computational time, we run simulations for only $7.6 \;\mathrm{Gyr}$; we do not assign a finite decay time when the vortex persists up to the end of our simulations, as in cases where $\msol/\mcentral \lesssim 1$.
The trend is as expected: vortices persist for shorter times when the soliton mass is large relative to the central mass.
There is some scatter in the decay times, so the curves are not perfectly monotonic.
This is due to the nonlinear nature of this scenario.

In \cref{fig:t_decay_vs_m_central}, we explore the relation between central mass and decay time, at fixed $\msol/\mcentral = 8$.
We see a negative correlation between $\mcentral$ and $\tdecay$.
This appears to be a power-law with $\tdecay \sim (80 \; \mathrm{Gyr}) {(\mcentral/10^7 \;\msolar)}^{-2.}$, but we caution against extrapolating outside this small mass range.
This can be explained by the increased density of solitons with higher mass. As the mass of the rotating soliton increases, the radius of the torus shrinks. So at the location of the soliton's maximum density, the soliton's self-gravity is a higher fraction of the background potential, leading to a greater influence of nonlinearities in the system's evolution and a shorter decay time.

\section{Discussion}
\label{sec:discussion}

Our calculations show that a background gravitational potential suppresses the decay of vortices in SDM\@.
Our simulations show that this suppression is sufficient to give vortices in soliton cores long lifetimes (compared e.g.\ to the dynamical/rotational time of the Milky Way, $\sim 0.25 \;\mathrm{Gyr}$) when the central gravitational field is generated by a black hole with a mass of order that of the soliton itself. The vortices are even longer-lived when the black-hole mass is greater than the soliton mass.
The lifetime of the vortices can exceed the Hubble time~\cite{Davies2020,Hui2017}. So, for all practical purposes, the vortices can be treated as stable, even though they are not the true lowest-energy state that carries angular momentum.

However, in an ultralight dark matter (ULDM) scenario of a $10^{-22}$ eV axion, using current estimates, supermassive black holes would typically be a few orders of magnitude too light to stabilize vortex-solitons in most galaxies. In the Milky Way, of halo mass $\sim 10^{12} M_\odot$, Sgr A* has a mass of $\sim 10^6 M_\odot$, while its soliton is estimated from simulations to be $\sim 10^9 M_\odot$\cite{Schive:2014dra}. For other supermassive black holes, given fiducial scalings of $M_{\rm core}\propto M_{\rm halo}^{1/3}$~\cite{Schive:2014dra} and $M_{\rm black~hole}\propto M_{\rm halo}^{1.55\pm 0.05}$~\cite{BoothSchaye2010}, we would expect black holes to typically achieve equal mass to soliton cores only for the most massive halos, with halo mass $\gtrsim 10^{15} M_\odot$, and black-hole and soliton masses of $\gtrsim 10^{10} M_\odot$. There are few halos thought to be as massive as $10^{15} M_\odot$; perhaps the largest cluster known, El Gordo, is thought to have mass $\sim 2\times 10^{15} M_\odot$ (each of two merging parts with mass $\sim 10^{15} M_\odot$). There are more than a handful of `ultramassive' ($M \ge 10^{10} M_\odot$) black holes known currently (e.g.\ 7 in \cite{2015ApJ...799..189Z}). As candidates for the most-massive,  Abell 1201 has been measured at $(3.27 \pm 2.12) \times 10^{10} M_\odot$~\cite{Nightingale:2023ini}, TON 618 has been measured at $\sim 4\times 10^{10} M_\odot$~\cite{GeEtal2019} and Phoenix A has mass perhaps $10^{11} M_\odot$~\cite{BrockampEtal2016}, even though a theoretical upper limit of $5\times 10^{10}$ has been estimated for a black hole accreting its mass through a disk~\cite{King2016}. Even the first-imaged black hole, M87*, has mass $(6.5\pm 1)\times 10^{9} M_\odot$~\cite{EventHorizon2019}, within striking distance of $10^{10}$; recall that the stabilization mechanism turns on gradually, still providing some stability when a black hole is lighter than the vortex-soliton. The assembly histories in these extreme clusters may be particularly complicated, but still, it seems worth considering whether these could have vortex-solitons. Several uncertainties need to be kept in mind, though, e.g.\ the scaling of soliton-core mass with halo mass has substantial uncertainty and scatter (e.g.\ \cite{Zagorac2022}). The black-hole mass scaling is uncertain as well; another estimate of the scaling exponent is 1.62 (\cite{MarascoEtal2021}, as used in Ref.\ \cite{PowellEtal2022}).

We originally conceived this stabilization mechanism for central ULDM solitons in galaxies, since ordinary matter and black holes would typically inhabit their centers as well. But it may be even more applicable to lower-mass solitons from higher-mass SDM particles. Due to the scale-invariance of the Gross--Pitaevskii--Poisson equations, the simulations in \cref{sec:simulations} solve the dynamics of a family of systems of characterized by the particle mass $m$.
The QCD axion has a much heavier particle mass, perhaps $m= \mathcal{O}(10^{-4}) \;\mathrm{eV}$, and may form solitons with mass on the order of $10^{-14}\,M_\odot$ and radius on the order of $300 \;\mathrm{km}$; the exact mass an radius depends on other properties of the axion~\cite{Sikivie2009,Levkov2018,Kirkpatrick2020,Kirkpatrick2022,Chavanis2021}. Our simulations suggest that in a gravitational well caused by ordinary hadronic matter of comparable mass, the soliton can support vortices in its cores with lifetimes of many dynamical times, going up to effectively infinite lifetime if the central mass is far-dominant. In an intermediate-mass scenario, a Solar-System-scale soliton may exist around the Sun. An axion mass of $\sim 10^{-14}$ eV would give an AU-scale vortex-less soliton, possibly detectable even if it is $\sim 12$ orders of magnitude less massive than the Sun~\cite{tsai2022direct}. 
A rotating version of this soliton seems quite plausible
and would have maximum density at $\sim 1 \;\mathrm{AU}$.

All of our analysis was with zero SDM self-interaction.
Previous work found that vortices are stable with repulsive self-interactions, but only when the mass of the soliton is \emph{larger} that some critical mass~\cite{Dmitriev:2021utv}.
Adding a central mass should reduce this critical mass.
On the other hand, we found that vortices without self-interactions are long-lived when the soliton mass is \emph{less} than a different critical mass, roughly equal to the central mass.
Adding self-interaction to our scenario would change this critical central mass too. Attractive self-interactions would destabilize the vortex, increasing the central mass threshold giving stability, and repulsive self-interactions would tend to stabilize the vortex, decreasing this threshold.

A full answer to the question of whether our gravitational route to SDM vortex stability actually enables stable spin-driven vortices in our Universe may require self-consistent cosmological simulations including SDM, hydrodynamics in the baryons, and black holes.
It also remains a question how a vortex-soliton around a black hole would form in the first place; it is plausible but not clear quantitatively when dynamical friction from rotating baryons would torque up SDM~\cite{Banik2013}. We think it possible that vortex-solitons naturally arise in a sufficiently fast-rotating halo, but have not shown an explicit formation mechanism or investigated how common the required level of rotation is in realistic halos.

In our study, we have treated the black hole as a nonrelativistic point mass potential, which is valid when the radius of the vortex-soliton is significantly longer than the Schwarzschild radius. When this assumption does not hold, a richer phenomenology can ensue. In this case, superradiance can convert a rotating black hole's spin to gravitational radiation and may generate a rotationally synchronized bosonic dark matter halo, with stability lifetimes possibly exceeding the age of the Universe~\cite{EastPretorius2017,HerdeiroRadu2017}. The gravitational stabilization we show here may contribute to the stability of such systems of `black holes with synchronized hair.' On the other hand, if a vortex-soliton surrounding a black hole has radius much larger than the black hole, the rotation would nearly evacuate the immediate surroundings of the black hole of dark matter, suppressing its interaction with the black hole.

Also, if black holes sometimes carry dark-matter vortex-solitons around with them, that is relevant to black-hole mergers and the role that dark matter plays in them~\cite{Bamber:2022pbs} including closing the `final parsec' of black-hole mergers~\cite{Milosavljevic:2002ht}.

In conclusion, we have demonstrated an alternative mechanism for the long-term stability of vortices in SDM:\@ a background gravitational potential suppresses their dominant decay mode.
We have primarily concerned ourselves with stabilization of vortex-solitons at the centers of dark matter halos comprised of particles with mass $\sim 10^{-22} \;\mathrm{eV}$ by supermassive black holes, but have highlighted other scenarios where this mechanism may be relevant. This introduces a new mechanism by which black holes and other pointlike masses might connect to their surroundings.

\section*{Author Contributions}

All authors contributed substantial ideas across the various concepts in the paper, but here we list particular contributions. NG contributed expertise with SDM simulations. AEM and NM did the bulk of the analysis and writing; AEM concentrated on the analytic energy arguments; NM wrote, designed, ran, and analysed simulations. MN initially conceived of the project that evolved into the current paper, and contributed large-scale-structure expertise and writing. CPW organized the working group, contributed ideas about SDM and its phenomenology from a particle-theory perspective, and guided the paper's and project's coherence.

\acknowledgments{}
\label{sec:acknowledgments}

We thank Luna Zagorac, Dmitry Levkov, David Mattingly and  Patrick Cheong for helpful discussions.
We would also like to thank the administrative and facilities staff at the University of New Hampshire including Katie Makem-Boucher and Michelle Mancini.

Computations were performed on Marvin, a Cray CS500 supercomputer at UNH supported by the NSF MRI program under grant AGS-1919310.
AEM's contributions to this project were supported by DOE Grant DE-SC0020220. NG's participation was supported in part by the National Science Foundation under Grant No. 1929080. MN acknowledges support by the Spanish grant PID2020-114035GB-100 (MINECO/AEI/FEDER, UE).
This work was initiated and performed in part at Aspen Center for Physics, which is supported by National Science Foundation under Grant No. PHY-1607611. CPW thanks the late Karsten Pohl for actively supporting the application for NSF grant No.\ 1929080.

This paper honors the memory of both Keenan Anderson and Tyre Nichols.

\bibliographystyle{JHEP}
\bibliography{bib}

\end{document}